\documentclass[intlimits,twoside,a4paper]{article}
\usepackage[cp1251]{inputenc}
\usepackage[eqsecnum]{cmpj3}

\newcommand{\be}{\begin{equation}}
\newcommand{\ee}{\end{equation}}
\newcommand{\bea}{\begin{eqnarray}}
\newcommand{\eea}{\end{eqnarray}}

\issue{2022}{25}{2}{23701}
\doinumber{10.5488/CMP.25.23701}

\title[Effect of hydrostatic pressure]%
{Effect of hydrostatic pressure on dynamic dielectric characteristics of CsH$_2$PO$_4$ ferroelectric}
\author[A. S. Vdovych, I. R. Zachek, R. R. Levitskii]{A. S. Vdovych\orcid{0000-0002-1888-8664}\refaddr{label1}\thanks{Corresponding author: vas@icmp.lviv.ua.}, R. R. Levitskii\refaddr{label1}, I. R. Zachek\refaddr{label2} }

\addresses{
	\addr{label1} Institute for Condensed Matter Physics of the National Academy of Sciences of Ukraine,\\
	1 Svientsitskii St., 79011 Lviv, Ukraine
	\addr{label2} Lviv Polytechnic National University, 12 Bandera Str., 79013 Lviv, Ukraine 
}

\date{Received March 15, 2022, in final form April 15, 2022}
\begin{document}
	
\maketitle

\begin{abstract}
Based on the pseudospin model of the deformed CsH$_2$PO$_4$ crystal  within the Glauber method,
the equation for the time-dependent mean value of the pseudospin is obtained, which is solved in the case of small deviations from the equilibrium state.
Using the solution of the equation, we find expressions for the longitudinal dynamic dielectric constant and relaxation time.
Based on the proposed parameters of the theory, the temperature and frequency dependences of the dynamic dielectric constant and the temperature dependence of the relaxation time are calculated and investigated. A detailed numerical analysis of the obtained results was performed. The influence of hydrostatic pressure on the dynamic characteristics of CsH$_2$PO$_4$ is studied.
%
%
%
\keywords ferroelectricity, ferroelectric phase transition, permittivity, mechanical deformation
%
\end{abstract}


\section{Introduction}

\begin{figure}[!t]
	\begin{center}
		\includegraphics[scale=0.6]{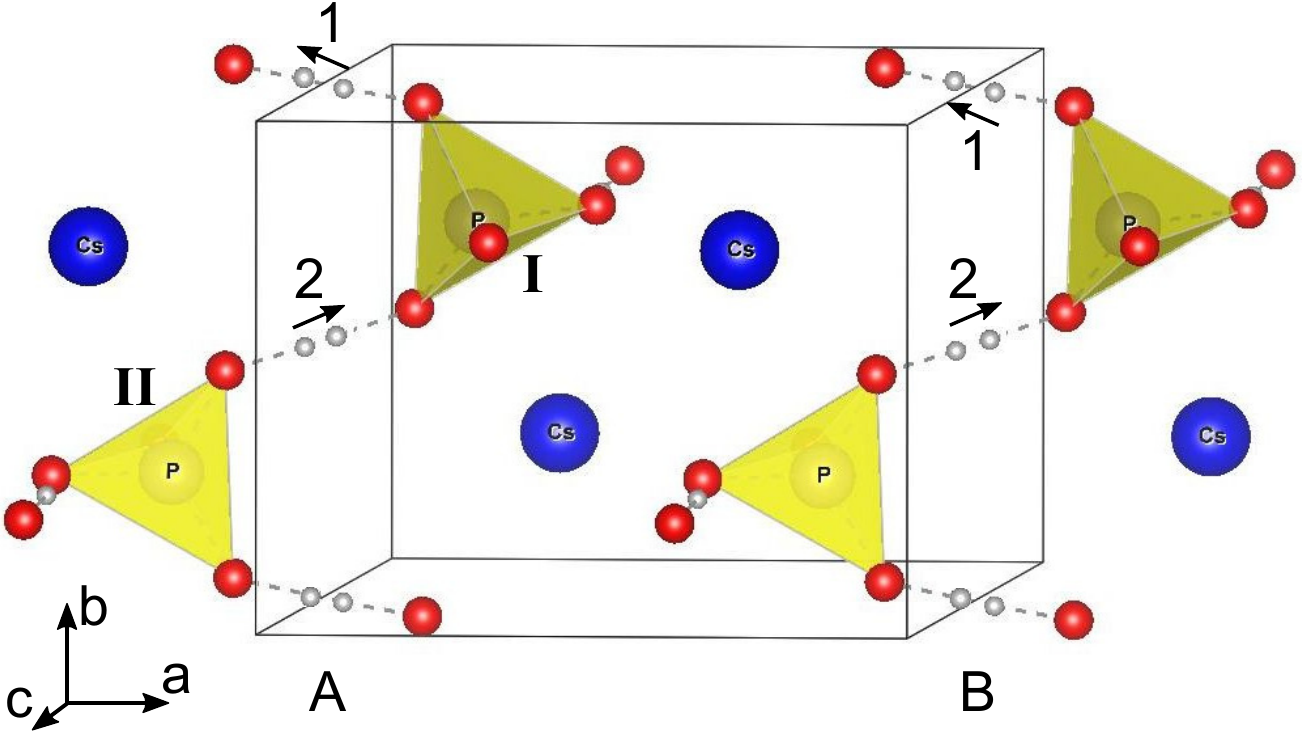}\\
		a\\
		~\\
		\includegraphics[scale=0.6]{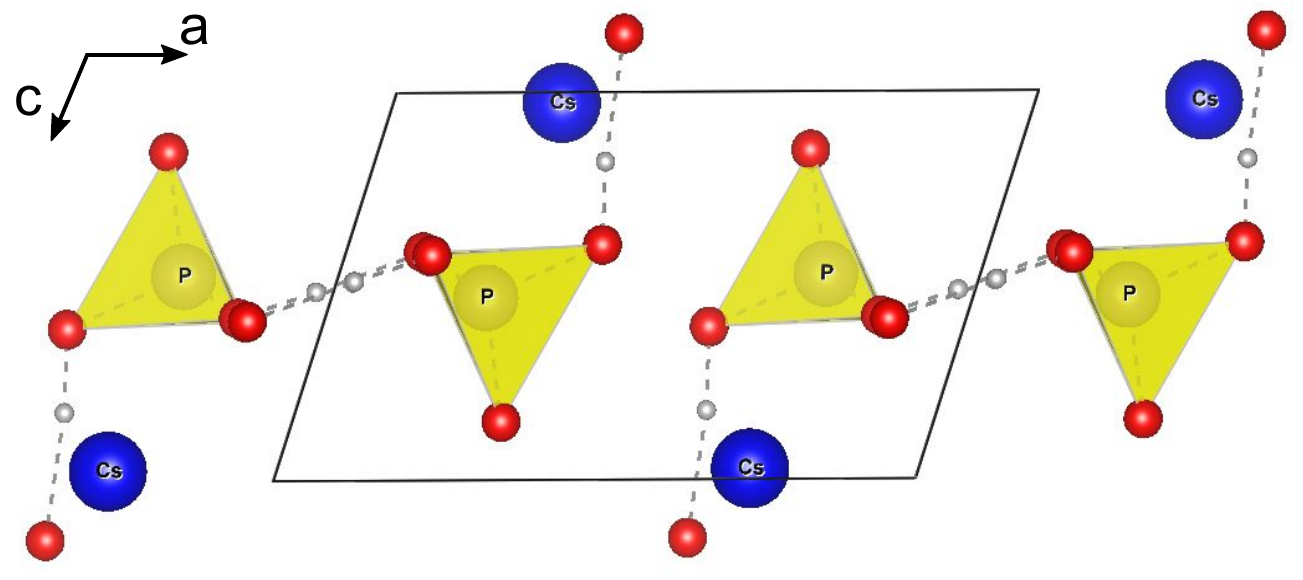}\\
		b
	\end{center}
	\caption{(Colour online) Primitive cell of CDP crystal in the ferroelectric phase \cite{Vdovych33702}.} \label{CDP_ferro_ab}
\end{figure}

A ferroelectric with hydrogen bonds CsH$_2$PO$_4$ (CDP) is an example of a crystal where the effect of pressure is significant. In this crystal there are two structurally equivalent types of hydrogen bonds of different lengths (figure~\ref{CDP_ferro_ab}b). Longer bonds have one equilibrium position for protons, and shorter bonds have two equilibrium positions. They connect the groups PO$_4$ in chains along the $ b $ -axis (figure~\ref{CDP_ferro_ab}a); therefore, the crystal is quasi-one-dimensional.

At room temperature in the absence of pressure, the crystal is in the paraelectric phase and has monoclinic symmetry (space group P2$_{1}$/m) \cite{Matsunaga2011,Itoh2626}. In this case, the protons on the short bonds are in two equilibrium positions with equal probability. Below $T_{c}=153$~K, the crystal passes into the ferroelectric phase (space group P2$_{1}$) \cite{Iwata304,Iwata4044} with spontaneous polarization along the crystallographic $b$ axis, and protons with a higher probability occupy the upper position (figure~\ref{CDP_ferro_ab}a).
Based on dielectric studies~\cite {Yasuda1311, Yasuda2755} it was established that at pressures $p=p_{c} =0.33$~GPa and $T_{c}^{\text{cr}}=124.6$~K, double hysteresis loops appear, i.e., there is a transition to the antiferroelectric phase. Using the neutron diffraction study \cite{Schuele935}, it was found that in the antiferroelectric phase, the unit cell of the CDP crystal doubles along the \textit{a}-axis, since there are two sublattices in the form of planes \textit{bc}, which are polarized antiparallel along $b$-axis and alternate along \textit{a}-axis. The symmetry remains monoclinic (space group P2$_{1}$).
Protons on hydrogen bonds are arranged in adjacent sublattices in an antiparallel manner.

The effect of hydrostatic pressure on the phase transition temperature and dielectric properties of ferroelectrics Cs(H$_{1-x}$D$_x)_2$PO$_4$ was studied in \cite{Gesi,Yasuda1311,Yasuda2755,Brandt,Kobayashi83,Magome2010,Deguchi024106}.
The molar heat capacity of CDP was measured in \cite{Imai3960}, and was also calculated on the basis of lattice dynamics modelling in \cite{Shchur054301,Shchur569}.
The important role of proton tunneling on bonds was shown based on the first-principle calculations \cite{Lasave134112} and using theoretical calculations on the basis of quasi-one-dimensional pseudospin model \cite{Kojyo4391}. 

Using \textit{ab initio} calculations,  piezoelectric coefficients, elastic constants, and molar heat capacity of CDP \cite{Shchur301,VanTroeye024112} were obtained.

The theoretical description of the dielectric properties of CDP at different values of hydrostatic pressure was carried out in \cite{Blinc6031,914R} on the basis of the pseudospin model.
However, in these works, the interaction parameters do not depend on the deformations of the lattice. As a result, it is impossible to obtain piezoelectric and elastic characteristics of the crystal, and the critical pressure does not depend on temperature.

In \cite{Deguchi3074}, the temperature dependences of lattice strains $u_1$, $u_2$, $u_3$, $u_5$ were measured. It also proposes a quasi-one-dimensional Ising model for the CDP crystal, in which the interaction parameters are linear functions of these strains. Based on this model, the temperature behavior of $u_j(T)$ was explained. However, this model does not consider the crystal as two sublattices and does not make it possible to describe the ferroelectric-antiferroelectric transition at high pressures.

In \cite{FXTT40,Levitskii4702,Vdovych33702,Vdovych_JPS2021_CDP}, there is proposed a two-sublattice pseudospin model of the deformed crystal CDP, in which the interactions between the nearest pseudospins in the chain are considered  in the mean field approximation. In this case, the interaction parameters are linear functions of the strains $u_j$. As a result, the temperature dependences of spontaneous polarization, dielectric constant, piezoelectric coefficients and elastic constants are calculated, and the influence of hydrostatic and uniaxial pressures and longitudinal electric field on these characteristics is studied.

In this paper, based on the model \cite{Vdovych33702}, the dynamic dielectric characteristics of CDP in the presence of hydrostatic pressure are calculated.

\section{The model of CDP crystal}

To calculate the dynamic characteristics of CDP, we use the model  \cite{Vdovych33702},
which considers a system of protons on O-H ... O bonds with two-minimum potential as a system of pseudospins.
The primitive cell has one chain, marked in figure \ref{CDP_ferro_ab} as ``A''. To describe the transition to antiferroelectric phase at high pressures, paper
\cite{Vdovych33702} considers an extended primitive cell formed by two chains (``A'' and ``B''). All chains ``A'' form a sublattice ``A'', and all chains ``B'' form a sublattice ``B''. Each chain in the primitive cell contains two adjacent tetrahedra PO$_4$ (type ``I'' and ``II'') together with two short hydrogen bonds (respectively, ``1'' and ``2'').
Dipole moments $\vec{d}_{q1}^\text{A}$, $\vec{d}_{q2}^\text{A}$, $\vec{d}_{q1}^\text{B}$, $\vec{d}_{q2}^\text{B}$, are ascribed to the protons on the bonds.  
Pseudospin variables
${\sigma_{q1}^\text{A}}/{2}$, ${\sigma_{q2}^\text{A}}/{2}$, ${\sigma_{q1}^\text{B}}/{2}$, ${\sigma_{q2}^\text{B}}/{2}$  describe  reorientation of the respective dipole moments of the base units: $\vec{d}_{q1,2}^\text{A,B} = \vec{\mu}_{q1,2}^\text{A,B} {\sigma_{q1,2}^\text{A,B}}/{2}$.

Under the stresses maintaining the symmetry of crystal $\sigma_1 = \sigma_{xx}$, $\sigma_2 = \sigma_{yy}$, $\sigma_3 = \sigma_{zz}$,  $\sigma_5 = \sigma_{xz}$ \linebreak($X$ $\perp$ (\textit{b},\textit{c}), $Y$ $\parallel$ \textit{b}, $Z$ $\parallel$ \textit{c}),  and in the presence of electric field  $E_2=E_y$ Hamiltonian of the model of CDP is given by \cite{Vdovych33702}:
\setcounter{equation}{0}
\renewcommand{\theequation}{2.\arabic{equation}}
\bea
&& \hat H= NU_{\text{seed}} + \hat H_{\text{short}} + \hat H_{\text{long}} + \hat H_{E} + \hat H'_{E}, \label{H_CDP}
\eea
where $N$  is the total number of restricted primitive cells.

The first term in (\ref{H_CDP})  is ``seed'' energy, which relates  to the heavy ion sublattice and does not explicitly depend on the configuration of the proton subsystem. 
It includes elastic, piezolectric and dielectric parts, expressed in terms of the electric field
$E_2$  and strains maintaining the symmetry of crystal $u_1=u_{xx}$, $u_2=u_{yy}$, $u_3=u_{zz}$, $u_5=2u_{xz}$:
\bea
&&  \hspace{-4ex} U_{\text{seed}} = v\left\lbrace\frac12 \sum\limits_{j,j'} c_{jj'}^{E0} u_ju_j'   - \sum\limits_{j} e_{2j}^0E_2u_j   - \frac12 \chi_{22}^{\varepsilon 0}E_2^2 \right\rbrace , ~~~~ j, j'=1,2,3,5,  \label{Useed}
\eea
Parameters  $c_{jj'}^{E0}$, $e_{2j}^0$, $\chi_{22}^{u 0}$ are the so-called ``seed'' elastic constants,  ``seed'' coefficients of
piezoelectric stresses and  ``seed'' dielectric susceptibility, respectively; $v$ is the volume of a restricted primitive cell. 
In the paraelectric phase, all coefficients $e_{ij}^0 \equiv 0$.

The other terms in  (\ref{H_CDP}) describe the pseudospin part of Hamiltonian.
In particular, the second term in  (\ref{H_CDP}) is Hamiltonian of short-range interactions:
\be
\hat H_{\text{short}} = - 2w \sum\limits_{qq'} \left ( \frac{\sigma_{q1}^\text{A}}{2} \frac{\sigma_{q'2}^\text{A}}{2} + \frac{\sigma_{q1}^\text{B}}{2}\frac{\sigma_{q'2}^\text{B}}{2} \right) \bigl( \delta_{{\vec R}_q{\vec R}_{q'}} + \delta_{{\vec R}_q + {\vec R}_b,{\vec R}_{q'}} \bigr). \label{Hshort}
\ee
In (\ref{Hshort}),  $\sigma_{q1,2}^\text{A,B}$ are  $z$-components of pseudospin operator that describe the state of the bond ``1''  or ``2'' of the chain ``A'' or ``B'', in the  $q$-th cell, ${\vec R}_b$ is the lattice vector along $OY$-axis.
The first Kronecker delta corresponds to the interaction between protons in the chains near the tetrahedra PO$_{4}$ of type ``I'', where the second  Kroneker delta is near the tetrahedra PO$_{4}$ of type ``II''. 
Contributions into the energy of interactions between  pseudospins near tetrahedra of different type are identical.
Parameter $w$,  which describes the short-range interactions within the chains, is expanded
linearly into a series with respect to strains $u_j$:
\be
w = w^0 + \sum\limits_{j} \delta_{j}u_j,(j=1,2,3,5). \label{w}
\ee

The  term $\hat H_{\text{long}}$ in  (\ref{H_CDP}) describes  long-range dipole-dipole interactions and indirect  (through the lattice vibrations)  interactions between protons which are taken into account in the mean field approximation:
\be
\hat H_{\text{long}} = N H^0 + \hat H_2, \label{Hlongg}
\ee
where such notations are used:
\bea
&&  \hspace{-8ex}\hat H^{0} = \nu_{1}( \eta_{1}^2 + \eta_{2}^2) + 2\nu_{2}\eta_{1}\eta_{2}, \\
&&  \hspace{-8ex} \hat H_{2} = \sum\limits_{q} \left\lbrace  - \left(2\nu_{1}\eta_{1} +  2 \nu_{2}\eta_{2} \right)   \! \left( \!\frac{\sigma_{q1}^\text{A}}{2} + \frac{\sigma_{q2}^\text{A}}{2} \!\right)\! - (2\nu_{2}\eta_{1} +  2 \nu_{1}\eta_{2} )  \! \left(\! \frac{\sigma_{q1}^\text{B}}{2} + \frac{\sigma_{q2}^\text{B}}{2} \!\right)\right\rbrace ,  \nonumber \\
&&  \hspace{-8ex} \nu_{1} = \nu_{1}^0 + \sum\limits_{j}  \psi_{j1}u_j,~  \nu_{2} = \nu_{2}^0 + \sum\limits_{j}  \psi_{j2}u_j, ~~~~ \langle \sigma_{q1}^\text{A} \rangle = \langle \sigma_{q2}^\text{A} \rangle = \eta_{1}, \quad
\langle \sigma_{q1}^\text{B} \rangle = \langle \sigma_{q2}^\text{B} \rangle = \eta_{2}. \label{nu}
\eea
The parameter $\nu_{1}$ describes the effective long-range interaction of the pseudospin that with the pseudospins inside the sublattice, and parameter $\nu_{2}$ - with the pseudospins of the other sublattice.

The fourth term in  (\ref{H_CDP}) describes the interactions of pseudospins with the external electric field:
\bea
&&\hat H_{E}= - \sum\limits_q \mu_y E_2 \left( \frac{\sigma_{q1}^\text{A}}{2} + \frac{\sigma_{q2}^\text{A}}{2} + \frac{\sigma_{q1}^\text{B}}{2} + \frac{\sigma_{q2}^\text{B}}{2} \right),
\eea
where $\mu_y$ is \textit{y}-component of effective dipole moments per one pseudospin.

The term $\hat H'_{E}$ in Hamiltonian (\ref{H_CDP}) takes into account the dependence of effective dipole moments on the mean value of pseudospin $s_f$:
\bea
&&\hat H'_{E} = -\sum\limits_{qf} s_f^2 \mu' E_2 \frac{\sigma_{qf}}{2} = - \sum\limits_{qf} \left(\frac{1}{N}\sum\limits_{q'}\sigma_{q'f}\right)^2 \mu' E_2 \frac{\sigma_{qf}}{2}, \label{H_E}
\eea
where $ \sigma_{qf} $ (\textit{f}=1, 2, 3, 4) are a brief notation of pseudospins $\sigma_{q1}^\text{A}$, $\sigma_{q2}^\text{A}$, $\sigma_{q1}^\text{B}$, $\sigma_{q2}^\text{B}$, respectively. 
Here, we use corrections to dipole moments $s_f^2 \mu'$ instead of $s_f \mu'$ because of the symmetry considerations, and the energy should not change when the field and all pseudospins change their signs.

The term $\hat H'_E$, as well as long-range interactions, are taken into account in the mean field approximation: 
\bea
&& \hspace{-8ex}  \hat H'_{E} =   -3 \sum\limits_{q} \mu'E_2 \left( \frac{\eta_1^2\sigma_{q1}^\text{A}}{2} + \frac{\eta_1^2\sigma_{q2}^\text{A}}{2} + \frac{\eta_2^2\sigma_{q1}^\text{B}}{2} + \frac{\eta_2^2\sigma_{q2}^\text{B}}{2} \right)  + 2N(\eta_1^3+\eta_2^3) \mu'E_2.
\eea

Our theory does not explicitly take into account the tunneling effects  because  the tunneling of protons on the bonds and strains of the lattice being taken into account simultaneously greatly complicates the calculations. In our model, the tunneling parameter is partially taken into account in the form of the other renormalized  parameters, such as the interaction parameter $w^0$ and the effective dipole moment $\mu_y$. Explicit consideration of tunneling could somewhat improve the agreement of the calculated characteristics with the experimental data.

In the two-particle cluster approximation for short-range interactions, the thermodynamic potential is given by:
\bea
&& G = N U_{\text{seed}} + NH^0  + 2N(\eta_1^3+\eta_2^3) \mu'E_2 - N v \sum\limits_{j}  \sigma_j u_j  \nonumber\\
&& - k_{\text B} T \sum\limits_q \left\{ 2 \ln {\rm Sp}\, \re^{-\beta \hat H^{(2)}_{q}}  \left. -  \ln {\rm Sp}\, \re^{-\beta \hat H^{(1)\text{A}}_{q}} -  \ln {\rm Sp}\, \re^{-\beta \hat H^{(1)\text{B}}_{q}} \right\} \right., \label{G}
\eea
where $\beta=\frac{1}{k_{\text B}T}$, $k_{\text B}$ is Boltzmann constant, $\hat H^{(2)}_{q}$, $\hat H^{(1)\text{A}}_{q}$, $\hat H^{(1)\text{B}}_{q}$ are two-particle and one-particle Hamiltonians:
\bea
&& \hspace{-6ex} \hat H^{(2)}_{q} = - 2 w\left( \frac{\sigma_{q1}^\text{A}}{2} \frac{\sigma_{q2}^\text{A}}{2} +
\frac{\sigma_{q1}^\text{B}}{2} \frac{\sigma_{q2}^\text{B}}{2}\right) -  \frac{y_{1}}{\beta} \left( \frac{\sigma_{q1}^\text{A}}{2} + \frac{\sigma_{q2}^\text{A}}{2} \right) -
\frac{y_{2}}{\beta} \left( \frac{\sigma_{q1}^\text{B}}{2} + \frac{\sigma_{q2}^\text{B}}{2} \right), \label{H2q}\\
&& \hspace{-6ex} \hat H^{(1)\text{A}}_{q} = - \frac{\bar y_{1}}{\beta} \left( \frac{\sigma_{q1}^\text{A}}{2} + \frac{\sigma_{q2}^\text{A}}{2} \right), ~~
\hat H^{(1)\text{B}}_{q} = - \frac{\bar y_{2}}{\beta} \left( \frac{\sigma_{q1}^\text{B}}{2} + \frac{\sigma_{q2}^\text{B}}{2} \right), \label{H1q}
\eea
where such notations are used:
\bea
&& y_{1} = \beta \Delta_{1} + 2\beta \nu_{1}\eta_{1}  +   2\beta \nu_{2} \eta_{2}+ \beta(\mu_yE_2 + 3\eta_1^2 \mu'E_2),  \label{y1}\\
&& y_{2} = \beta \Delta_{2} + 2\beta \nu_{2}\eta_{1}  +   2\beta \nu_{1} \eta_{2}+ \beta(\mu_yE_2 + 3\eta_2^2 \mu'E_2), \nonumber\\
&& \bar y_{1} =  \beta \Delta_{1} + y_{1}, ~~ \bar y_{2} =  \beta \Delta_{2} + y_{2}. \nonumber
\eea
Symbols $\Delta_l$  are the effective fields created by the neighboring bonds
from outside the cluster. 
From the condition of the minimum of the thermodynamic potential $\partial G/\partial \Delta_l =0$, the system of equations for these fields, as well as for the order parameters $\eta_l$ are obtained:
\bea
&& \eta_{1} = \frac{1}{D} \left[ \sinh (y_{1} + y_{2}) +  \sinh (y_{1} - y_{2}) + 2 a\sinh y_{1}\right] = \tanh \frac{\bar y_{1}}{2}, \nonumber\\
&& \eta_{2} = \frac{1}{D} \left[ \sinh (y_{1} + y_{2}) -  \sinh (y_{1} - y_{2}) + 2 a \sinh y_{2} \right] = \tanh \frac{\bar y_{2}}{2}, 
\label{eta1eta2}
\eea
Here, the following notations are used:
\bea
&& D = \cosh (y_{1} + y_{2}) +  \cosh (y_{1} - y_{2}) + 2a \cosh y_{1}+ 2a \cosh y_{2}  + 2a^2,  ~~~~ a = \re^{-\frac{w}{k_{\text B}T}}.\nonumber
\eea

Minimizing the thermodynamic potential with respect to the strains, an additional system of equations for the strains $u_j$ was obtained in \cite{Vdovych33702}:
\bea
 \sigma_j &=& c_{j1}^{E0}u_1 + c_{j2}^{E0}u_2 + c_{j3}^{E0}u_3 + c_{j5}^{E0}u_5 - e_{2j}^0E_2 - \frac{2\delta_j}{v} + \frac{4\delta_j}{vD}M \nonumber   \\
& -& \frac{1}{v} \psi_{j1} ( \eta_{1}^{2}  +  \eta_{2}^{2}  ) - \frac{2}{v} \psi_{j2} \eta_{1}\eta_{2}. 
\label{sigma}
\eea
where
\[ M=\bigl[   a\,\cosh y_{1} + a\,\cosh y_{2} + 2\,a^{2} \bigr]. \]

In the case of ferroelectric ordering  $\eta_{1} = \eta_{2} = \eta$,  $y_{l}=y_{2}=y$. Then, (\ref{y1}), (\ref{eta1eta2}) are given by:
\bea
&& y = \beta \Delta + 2\beta (\nu_{1}+\nu_{2})\eta  + \beta(\mu_yE_2 + 3\eta_1^2 \mu'E_2), ~~~~ \bar y =  \beta \Delta + y, \label{y} 
\eea
\bea
&&  \hspace{-4ex}\eta = \frac{ \sinh (2y) +  2a \sinh y}
{\cosh (2y) + 1 + 4a \cosh y  + 2a^2}  = \tanh \frac{\bar y}{2}  =  \frac{ \sinh y \cosh y +  a \sinh y }
{\cosh^2 y + 2a \cosh y  + a^2}  =  \frac{ \sinh y} 
{\cosh y + a}. \label{eta}
\eea
In the presence of hydrostatic pressure $\sigma_1=\sigma_2=\sigma_3=-p$,  $\sigma_4=\sigma_5=\sigma_6=0$.
In \cite{Vdovych33702}, the expression for the longitudinal component of polarization $P_2$ is also obtained, which in the case of ferroelectric ordering is given by:
\bea
&& \hspace{-4ex} P_2 =  \sum\limits_{j} e_{2j}^0u_j   + \chi_{22}^{u 0}E_2  + \frac{2\mu_y}{v} \eta  +  \frac{2\mu'}{v} \eta^3. \label{P2}
\eea

\section{Dynamic dielectric properties of mechanically clamped CDP crystal. Analytical results}

Dynamic dielectric properties are studied in the absence of the electric field and shear stresses.
To study the dynamic properties of the CDP crystal, an approach is used which is
based on the ideas of Glauber's stochastic model. Based on this approach, we obtain a system of equations for time-dependent unary correlation functions of pseudospins:
\setcounter{equation}{0}
\renewcommand{\theequation}{3.\arabic{equation}}
\be
- \alpha \frac{\rd}{\rd t} \langle  \sigma_{qf} \rangle   =   \left\langle \sigma_{qf} \left[ 1  -
\sigma_{qf} \tanh \frac12 \beta {\varepsilon}_{qf}(t)\right]  \right\rangle , \label{glauber_unar}
\ee
where the parameter $\alpha$ sets the time scale of the dynamic processes in the sytem; $\varepsilon_{qf'}(t)$ is the local field acting on the  $f'$-th pseudospin in the  $q$-th cell; this is the factor at $\sigma_{qf'}/2$ in the initial Hamiltonian.
In order to obtain a closed system of equations for unary correlation functions, we use the two-particle cluster approximation. In this approximation, the local fields $\varepsilon_{qf}(t)$ are the coefficients at $\sigma_{qf}/2$ in two-particle and one-particle Hamiltonians (\ref{H1q}), (\ref{H2q}). In the two-particle approximation, they are given by:
\bea
&& \varepsilon_{q1}  =   w\sigma_{q2}  +  \frac{y}{\beta}, \quad \varepsilon_{q2}  =   w\sigma_{q1}  +  \frac{y}{\beta}, \label{eps_f}
\eea
and in the two-particle approximation they are given by:
\bea
&& \varepsilon_{qf}  =  \frac{\bar{y}}{\beta}.
\eea
Here, the effective fields $y$, $\bar{y}$ are given by expressions (\ref{y1}).

As a result of (\ref{glauber_unar}), we obtain a system of equations for the mean values of pseudospins $ \langle \sigma_{qf} \rangle =\eta_{f}$ in the two-particle approximation
\bea
&& \alpha \frac{\rd}{\rd t} \eta  =  -(1-P) \eta  +  L,  \label{deta2} 
\eea
and in the one-particle approximation
\bea
&& \alpha \frac{\rd}{\rd t} \eta = - \eta + \tanh \frac{\bar y}{2}, \label{deta1}
\eea
where the following notations are used:
\bea
&& \!\!\!\!P  =  \frac12 \left[ \tanh \left( \frac{\beta w}{2} \! + \! \frac{y}{2} \right)  \!-\!  \tanh \left(- \frac{\beta w}{2} \! + \! \frac{y}{2} \right) \right],  ~~  L  =  \frac12 \left[ \tanh \left( \frac{\beta w}{2} \! + \! \frac{y}{2} \right)  \!+\!  \tanh \left(- \frac{\beta w}{2} \! + \! \frac{y}{2} \right) \right].\nonumber
\eea

We limit ourselves to solving the equations (\ref{deta2}) and (\ref{deta1}) to the case of small deviations from the equilibrium state. To do this, we present  $\eta_{f}$ and the effective fields $y_{f}$, $\bar y_{f}$ as the sum of two terms -- equilibrium values and their deviations from the equilibrium state (mechanically clamped crystal):
\bea
&& \hspace{-4ex} \eta = \tilde{\eta} + \eta_{t}, \nonumber\\
&& \hspace{-4ex}  y = \tilde{y} + y_{t} =   \beta [\Delta + 2(\nu_{1}+\nu_{2})\tilde{\eta} +  \Delta_{t} + 2(\nu_{1}+\nu_{2})\tilde{\eta}_{t}     +  (\mu_y + 3\tilde{\eta}_1^2 \mu')E_{2t}], ~~E_{2t} = E_2 \re^{\ri\omega t}.
\label{eta0t}
\eea
Here, $\Delta$ is the effective cluster field, and $\Delta_{t}$ is the deviation from its equilibrium value, $\nu_{i}$ is the parameters of long-range interactions.

We expand the coefficients $P_{f}$ and $L_{f}$ in series over $y_{ft}/{2}$ limited to linear terms:
\bea
&&P = P^{(0)} + \frac{y_{t}}{2} P^{(1)}, ~L = L^{(0)} + \frac{y_{t}}{2} L^{(1)}, ~
\tanh \frac{\bar y}{2} = \tanh \frac{\tilde{\bar y}}{2} + \varphi \frac{\bar y_{t}}{2}, \label{P0t}
\eea
where the following notations are used:
\bea
&&  \hspace{-4ex} P^{(0)}  = \frac{1-a^2}{Z}, ~~ P^{(1)}   =   \frac{-4a(1 - a^2)\sinh \tilde{y}}{Z^{2}}, ~~~~~  L^{(0)} = \frac{2a \sinh \tilde{y}}{Z}, ~ L^{(1)}   =   \frac{4a[2a   +   (1  +  a^2) \cosh \tilde{y}]}{Z^{2}}, \nonumber \\
&& Z=1 + a^2  +  2a \cosh \tilde{y}; ~~~~ \varphi = (1-\tilde{\eta}^2), ~~ a = \re^{-\frac{w}{k_{\text B}T}}, ~~ w = w^0 + \sum\limits_{j}\delta_ju_j,  ~~j=1,2,3,5.\nonumber
\eea

Substituting (\ref{eta0t}), (\ref{P0t}) in (\ref{deta2}), (\ref{deta1}), we obtain a system of differential equations for unary distribution functions:
\bea
&& \alpha \frac{\rd}{\rd t} \eta_{t}  =  -\left( 1-P^{(0)} - \frac{y_{t}}{2} P^{(1)}\right)  (\tilde{\eta} + \eta_{t})  +  L^{(0)} + \frac{y_{t}}{2} L^{(1)},\nonumber   \\
&& \alpha \frac{\rd}{\rd t} \eta_{t} =  - (\tilde{\eta} + \eta_{t}) + \tanh \frac{\tilde{\bar y}}{2} + \varphi \frac{\bar y_{t}}{2}. \label{deta} \eea
Since the equilibrium state holds  (\ref{eta}), it is easy to see that $(1-P^{(0)})\tilde{\eta}=L^{(0)}$. Then, the system of equations (\ref{deta}) is given by
\bea
&& \alpha \frac{\rd}{\rd t} \eta_{t}  =  -\left( 1-P^{(0)}\right)  \eta_{t}  + Y \frac{y_{t}}{2},   ~~~~~~ Y = \left( P^{(1)} \tilde{\eta} + L^{(1)}\right),  \nonumber  \\
&& \alpha \frac{\rd}{\rd t} \eta_{t} =  - \eta_{t} + \varphi \frac{\bar y_{t}}{2}.\label{detax}
\eea
Let us write it  more in detail, taking into account (\ref{y}):
\bea
&& \alpha \frac{\rd}{\rd t} \eta_{t}  =  -\left( 1-P^{(0)}\right)  \eta_{t}  + \frac{1}{2} \beta \left[ \Delta_{t} + 2(\nu_{1}+\nu_{2})\tilde{\eta}_{t}     +  (\mu_y + 3\tilde{\eta}^2 \mu')E_{2t}\right]  Y, \nonumber  \\
&& \alpha \frac{\rd}{\rd t} \eta_{t} =  - \eta_{t} + \varphi \frac{1}{2} \beta \left[ 2\Delta_{t} + 2(\nu_{1}+\nu_{2})\tilde{\eta}_{t}     +  (\mu_y + 3\tilde{\eta}^2 \mu')E_{2t}\right] .
\label{detaxx}
\eea
Excluding the parameter $\Delta_{ft}$, we obtain the equation for the time-dependent order parameter:
\bea
&&  \frac{\rd}{\rd t} \eta_{t}  =  A \eta_{t}  + B \beta (\mu_y + 3\tilde{\eta}^2 \mu')E_{2t} ,  \label{detat}
\eea
where
\bea
&& \hspace{-4ex}  A = \left[  -\frac{(1-P^{(0)})2\varphi}{Y}+1 + \varphi \beta(\nu_{1}+\nu_{2})\right]   \Big/ \left[  \alpha \left( \frac{2\varphi}{Y}-1\right) \right]  , ~~~~ 
 B = \frac{1}{2} \varphi \Big/ \left[ \alpha \left( \frac{2\varphi}{Y}-1\right) \right] .
\eea
Solving this equation, we obtain the time-dependent mean values of pseudospins, and thus we find
longitudinal dynamic dielectric susceptibility of mechanically clamped CDP crystal:
\bea
&&  \chi_{22}(\omega) = \chi_{22}^{0} + \lim\limits_{E_{2t}\to 0} \frac{2}{v} (\mu_y + 3\tilde{\eta}^2 \mu') \frac{\rd \eta_{t}}{\rd E_{2t}} =
\chi_{22}^{0} +  \frac{\chi}{1 + \ri\omega \tau}, \label{X22}
\eea
where $\chi$ is the static susceptibility of the pseudo-spin subsystem:
\bea
&& \hspace{-4ex} \chi = \frac{2\beta (\mu_y + 3\tilde{\eta}^2 \mu')^2}{v} \tau B,  \label{Xi}\nonumber
\eea
$\tau$ is relaxation time:
\be
\tau = - 1/A.\label{tau}
\ee
Dynamic dielectric constant CDP is given by:
\bea
&& \varepsilon_{22}(\omega) = 1+4\piup\chi_{22}(\omega). \label{eps22_om}
\eea

\section{Comparison of theoretical results with experimental data. Discussion of the obtained results}

The theory parameters are  determined in \cite{Vdovych33702} from the condition of agreement of calculated characteristics with experimental data for temperature dependences of spontaneous polarization $P_2(T)$ and dielectric permittivity  $\varepsilon_{22}(T)$ at different values of hydrostatic pressure  \cite{Yasuda2755}, spontaneous strains $u_j$ \cite{Deguchi3074}, molar heat capacity \cite{Imai3960} and elastic constants  \cite{Prawer63};  as well as the agreement with  ab initio calculations of the  lattice contributions to molar heat capacity  \cite{Shchur301} and dielectric permittivity at zero temperature \cite{VanTroeye024112}.

It should be noted that the temperature dependences of the dielectric constant $\varepsilon_{22}$ at different values  of hydrostatic pressure were also measured in \cite{Deguchi024106}. However, they do not agree with experimental data \cite{Yasuda2755}. It is quite possible that another crystal sample was used there, which was grown under different conditions. In addition, in \cite{Deguchi024106} there are no data for the temperature dependences of spontaneous polarization at different pressures, as well as there are no data for dielectric characteristics at zero pressure. Therefore, we used experimental data \cite{Yasuda2755} to determine the model parameters.
To describe the dynamic dielectric permittivity, we used the data \cite{Deguchi3575} because they are measured at frequencies corresponding to the  range of dispersion. In the case of \cite{Deguchi024106}, the frequencies were much lower (up to 1MHz). At these frequencies, the dielectric constant in the paraelectric phase behaves as static, and in the ferroelectric phase there is a large contribution to the permittivity associated with the reorientation of the domain walls. However, our theory does not take into account the reorientation of the domain walls and cannot describe the dielectric permittivity in the ferroelectric phase at low frequencies.

Parameters of short-range interactions $w_0$ and long-range interactions  $\nu_1^{0}$ (``intra-sublattice''), $\nu_2^{0}$ (``inter-sublattice'') mainly fix the phase transition temperature from  paraelectric to ferroelectric phase at the absence of external pressure and field, the order of  phase transition and the shape of curve $P_2(T)$. Their optimal values are: $w_0/k_\text{B}=650$~K,  $\nu_1^{0}/k_\text{B}=1.50$~K,  $\nu_2^{0}/k_\text{B}=0.23$~K.

To determine deformational potentials $\delta_{j}$ [see (\ref{w})] and $\psi_{j1}$ (\ref{nu}), $\psi_{j2}$ (\ref{nu}), it is necessary to use experimental data for the shift of the phase transition temperature under hydrostatic and uniaxial pressures as well as the data for temperature dependences of  spontaneous strains $u_j$, piezoelectric coefficients and elastic constants. Unfortunately, only the  data for the spontaneous strains and  hydrostatic pressure effect on the dielectric characteristics are available. 
As a result,  the experimental data for strains and dielectric characteristics can be described  using a great number of  combinations of parameters $\psi_{j1}$,  $\psi_{j2}$. 
Therefore, for the sake of simplicity, we chose $\psi_{j2}$ to be proportional to $\psi_{j1}$.  
Optimal values of deformational potentials are: 
$\delta_{1}/k_\text{B}=1214$~K, $\delta_{2}/k_\text{B}=454$~K, $\delta_{3}/k_\text{B}=1728$~K, $\delta_{5}/k_\text{B}=-131$~K;
$\psi_{11}/k_\text{B} = 92.2$~K,  $\psi_{21}/k_\text{B} = 23.2$~K,  $\psi_{31}/k_\text{B} =139.7$~K,  $\psi_{51}/k_\text{B} = 5.5$~K;
$\psi_{j2}$ = $\frac{1}{3}\psi_{j1}$.

The effective dipole  moment in the paraelectric phase is found  from the condition of agreement of the calculated curve $\varepsilon_{22}(T)$  with experimental data. We consider it to be dependent on the value of hydrostatic pressure \textit{p}, that is $\mu_{y}=\mu_{y}^0(1-k_pp)$, where $\mu_{y}^0=2.63\cdot 10^{-18}$esu$\cdot$cm, ~~$k_p=0.4\cdot 10^{-10}$cm$^2$/dyn. 
The correction to the effective dipole moment $\mu'=-0.43\cdot 10^{-18}$esu$\cdot$cm is found  from the condition of the agreement of calculated saturation polarization with experimental data.
In \cite{Deguchi3575}, the experimental data for the dielectric permittivity $\varepsilon_{22}$ are almost twice as large as in \cite{Yasuda2755}. This may be due to different growing conditions of the samples and different features of measurement the dielectric permittivity. Therefore, to describe the experimental data \cite{Deguchi3575} for $\varepsilon_{22}$, we assume that these samples are characterized by a more effective dipole moment per pseudospin. Namely, $\mu_{y}^0=5.2\cdot 10^{-18}$esu$\cdot$cm, $\mu'=-1.8\cdot 10^{-18}$esu$\cdot$cm.

The ``seed'' dielectric susceptibility  $\chi_{22}^{u 0}$, coefficients of piezoelectric stress $e_{2j}^0$ and elastic constants $c_{ij}^{E0}$  are found  from the condition of the agreement of theory  with experimental data in the temperature regions far from the phase transition temperature  $T_c$. Their values are obtained as follows:
$\chi_{22}^{u 0}=0.443$;  ~ $e_{2j}^0=0.0$~esu/cm$^2$; ~~
$c_{11}^{0E} = 28.83 \cdot 10^{10}$dyn/cm$^2$, ~$c_{12}^{E0} = 11.4 \cdot 10^{10}$dyn/cm$^2$, ~$c_{13}^{E0} = 42.87 \cdot 10^{10}$dyn/cm$^2$,~
$c_{22}^{E0} = 26.67 \cdot 10^{10}$dyn/cm$^2$,~
$c_{23}^{E0} = 14.5 \cdot 10^{10}$dyn/cm$^2$, ~$c_{33}^{E0} = 65.45 \cdot 10^{10}$dyn/cm$^2$, ~$c_{15}^{E0} = 5.13 \cdot 10^{10}$dyn/cm$^2$,  ~$c_{25}^{E0} = 8.4 \cdot 10^{10}$dyn/cm$^2$, ~$c_{35}^{E0} = 7.50 \cdot 10^{10}$dyn/cm$^2$,~
$c_{55}^{E0} = 5.20 \cdot 10^{10}$dyn/cm$^2$.

The volume of restricted primitive cell is  $\upsilon$ = 0.467$\cdot 10^{-21}$ cm$^3$ \cite{Schuele935}.

\begin{figure}[!t]
	\begin{center}
		\includegraphics[scale=0.85]{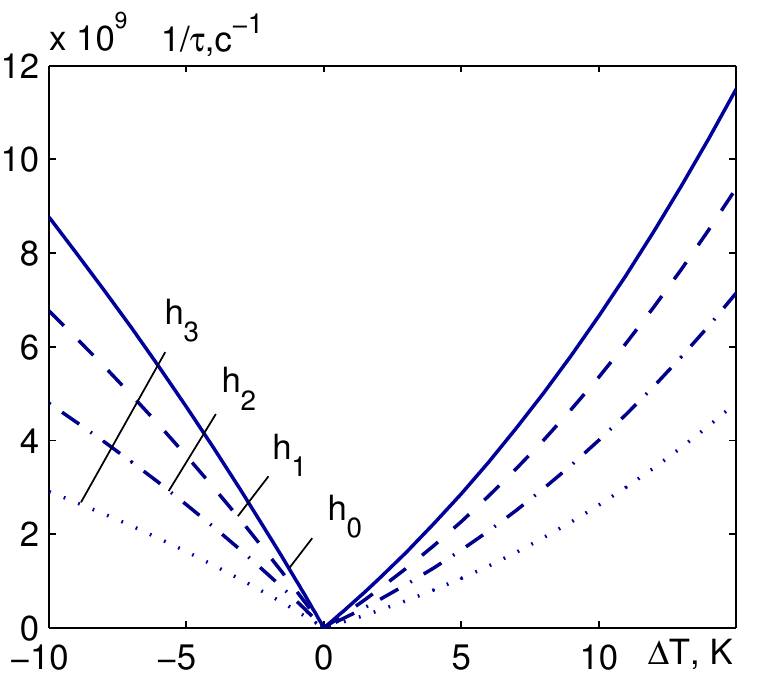} \includegraphics[scale=0.85]{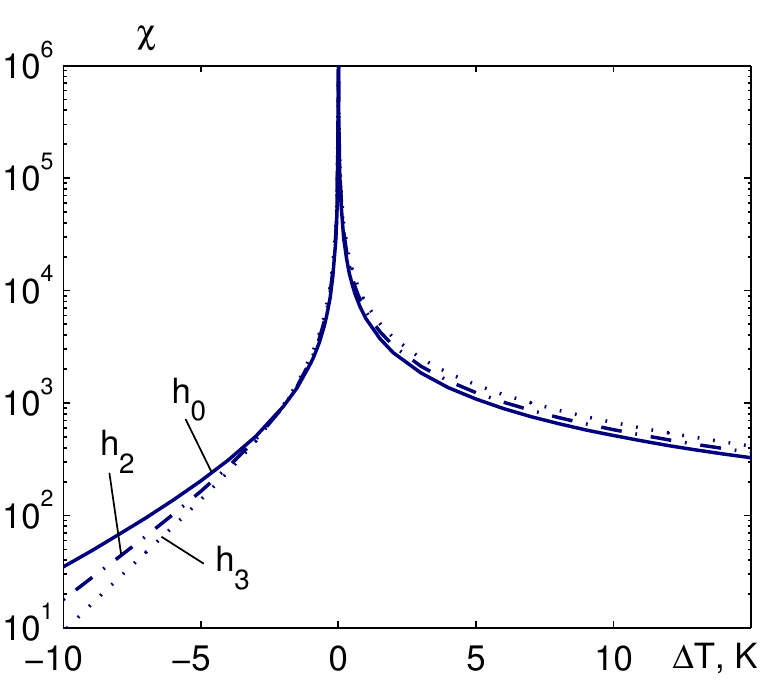}\\
		a  ~~~~~~~~~~~~~~~~~~~~~~~~~~~~~~~~~~~~~~~~~~~~~~~~~~~~~~~~~ b
	\end{center}
	\caption{(Colour online) The temperature dependence of the inverse relaxation time $\tau^{-1}$ and static dielectric susceptibility $\chi$ of the pseudospin subsystem  at different values of hydrostatic pressure.   In the notation $h_p$, the lower index means the value of pressure  $p$ (kbar).} \label{tau_p}
\end{figure}

As can be seen from (\ref{X22}), the dynamic dielectric susceptibility is determined by the behavior of the static dielectric susceptibility $\chi$ of the pseudospin subsystem and the relaxation time $\tau$ in the system.
Their temperature dependences at different values of pressure are shown in figure~\ref{tau_p}.

The calculated relaxation time $\tau$ tends to infinity at $T=T_c$.
It is associated with the relaxation frequency characteristic of this crystal (soft relaxation mode)
$\nu_{s}= (2\piup\tau)^{-1}$, which conditionally separates the region of low-frequency and high-frequency dispersion.
The inverse time $\tau^{-1}$, as well as  the relaxation frequency $\nu_{s}$ decrease while approaching the phase transition temperature and
tending to zero at a temperature $T=T_c$.

\begin{figure}[!t]
	\begin{center}
		\includegraphics[scale=0.85]{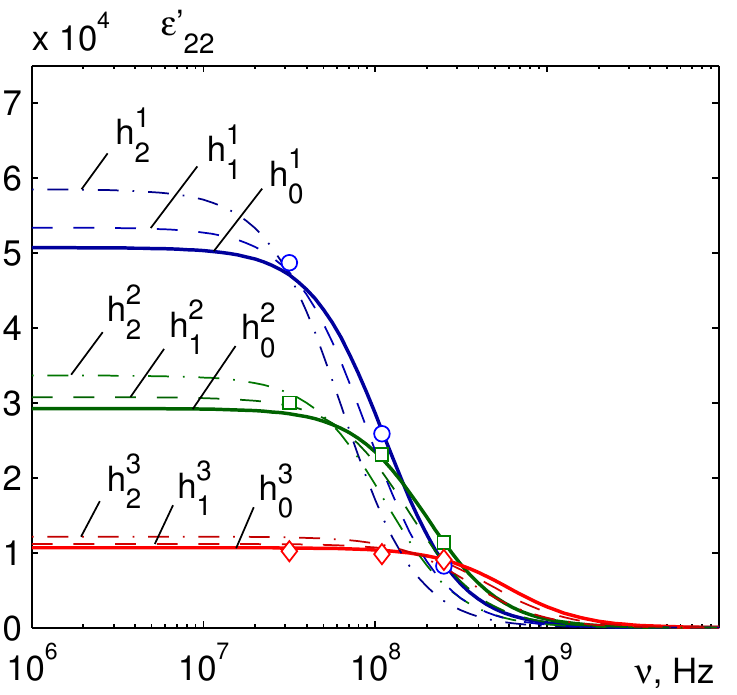}\includegraphics[scale=0.85]{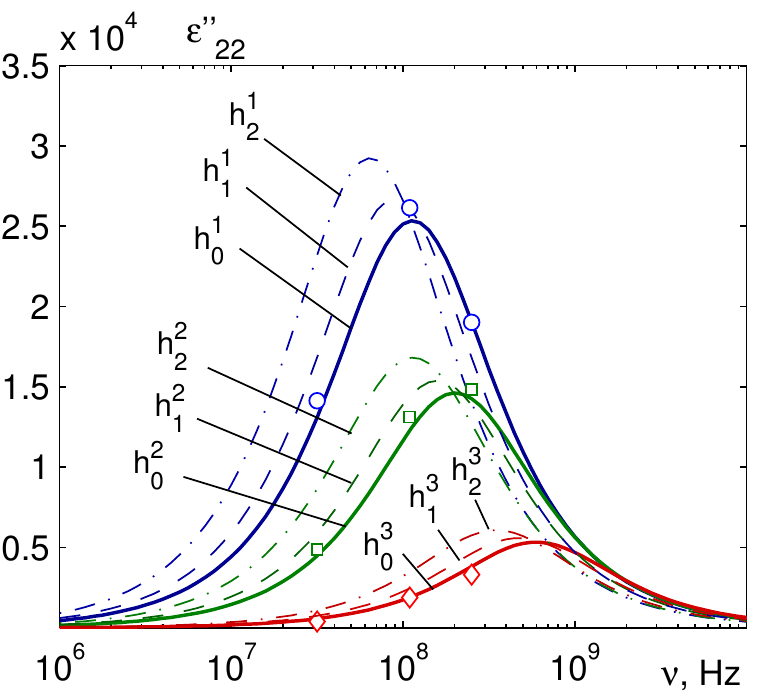}
	\end{center}
	\caption{(Colour online) The frequency dependences of the real $\varepsilon'_{22}$ and imaginary $\varepsilon''_{22}$ parts of the dielectric constant of CDP at different $\Delta T$ in paraelectric phase and at different values of hydrostatic pressure $p$. In the notation $h^{i}_p$, the superscripts number the temperature: $\Delta T=1.4$~K, $\circ$(1);    $\Delta T=2.4$ K, $\square$(2);  $\Delta T=6.3$ K, $\triangledown$(3),
		and the lower indices indicate the value of pressure (kbar).
		The symbols~$\circ$, $\square$, $\diamond$ are experimental data \cite{Deguchi3575}.} \label{e22re_nu_h}
\end{figure}

\begin{figure}[!t]
	\begin{center}
		\includegraphics[scale=0.85]{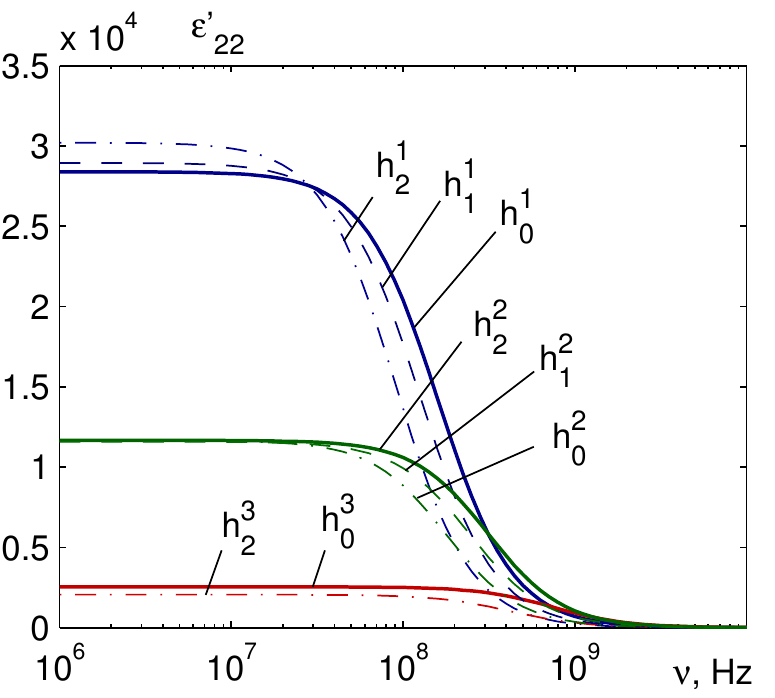}\includegraphics[scale=0.85]{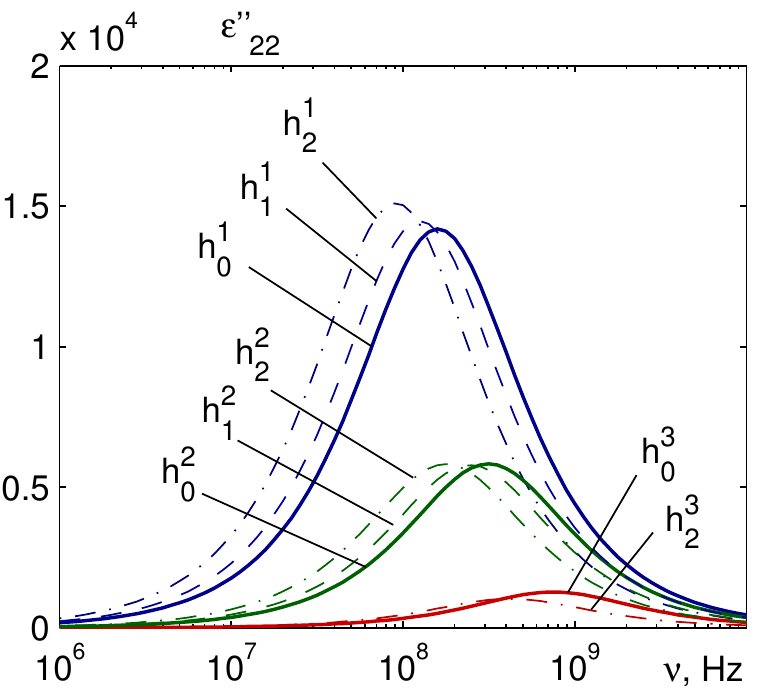}
	\end{center}
	\caption{(Colour online) The frequency dependences of the real $\varepsilon'_{22}$ and imaginary $\varepsilon''_{22}$ parts of the dielectric constant of CDP at different $\Delta T$ in ferroelectric phase and at different values of hydrostatic pressure $p$. In the notation $h^{i}_p$, the superscripts number the temperature: $\Delta T= -1$~K (1);   $\Delta T= -2$~K(2);  $\Delta T= -5$~K(3),
		and the lower indices indicate the value of pressure (kbar).} \label{e22re_nu_hs}
\end{figure}

\begin{figure}[!t]
	\begin{center}
		\includegraphics[scale=0.85]{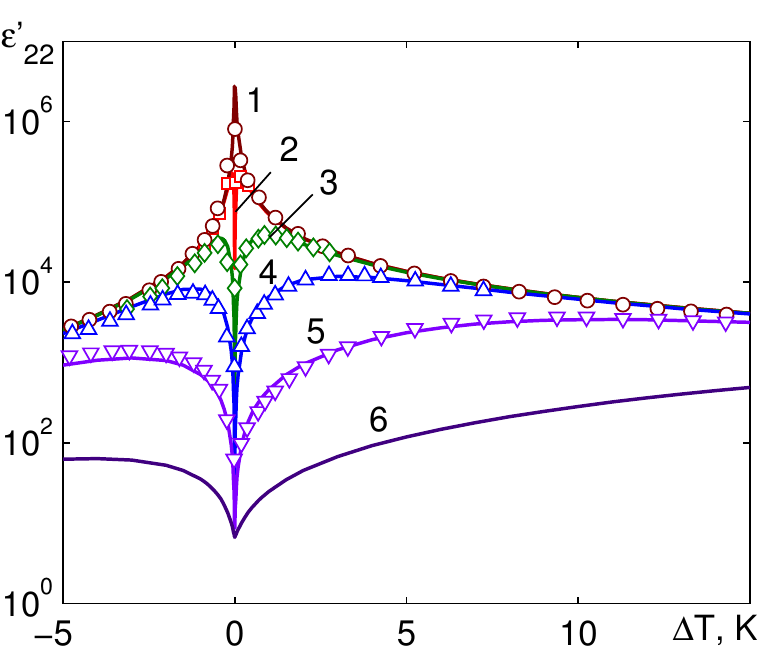}\includegraphics[scale=0.85]{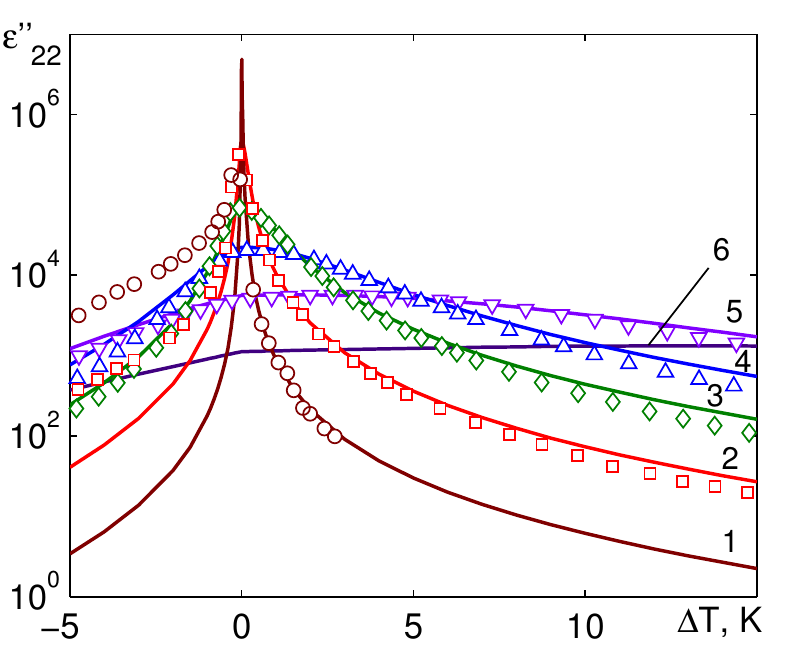}
	\end{center}
	\caption{(Colour online) The temperature dependences of  $\varepsilon'_{22}$ and $\varepsilon''_{22}$ of CDP
		at different frequencies  $\nu$ (MHz): 1.0 -- 1,$\circ$,\cite{Deguchi3575}; 12.0 -- 2,$\square$,\cite{Deguchi3575}; 72.4 -- 3, $\diamond$, \cite{Deguchi3575};  251.2 -- 4, $\vartriangle$, \cite{Deguchi3575};  1000.0 -- 5, $\triangledown$, \cite{Deguchi3575};  5000.0 -- 6.} \label{e22reT}
\end{figure}

At frequencies $\nu\ll\nu_{s}$, the real part of the dynamic dielectric constant $\varepsilon_{22}'$ behaves as static, and the imaginary part $\varepsilon_{22}'' $ is close to zero at all temperatures, except for a narrow area near $T_c$. This can be seen in the frequency dependences $\varepsilon_ {22}(\nu)$ for different $\Delta T = T-T_c$ in the frequency range $\nu<10^7$~Hz (figure~\ref{e22re_nu_h}, \ref{e22re_nu_hs}), as well as on the temperature dependences $\varepsilon_{22}(T)$ at low frequencies and temperatures far from $T_c$ (figure~\ref{e22reT}).

At frequencies $\nu \approx \nu_{s}$, there is a relaxation dispersion, which is manifested in a rapid decrease of the real part of the dielectric constant $\varepsilon_{22}'$ with increasing  frequency, as well as in large values of the imaginary part $\varepsilon_{22}''$; in which case the maximum value of $\varepsilon_{22}''$ corresponds to the frequency $\nu_{s}$. This can be seen in the frequency dependences $\varepsilon_{22} (\nu)$ for different $\Delta T = T-T_c$ in the frequency range $10^7<\nu<10^{10}$~Hz (figure~\ref{e22re_nu_h},\ref{e22re_nu_hs}). At the temperature dependences $\varepsilon_{22} (T)$, the relaxation dispersion is manifested in the decrease of $\varepsilon_{22}'$ near $T_c$ in the form of a depression, as well as in the presence of the peak $\varepsilon_{22}''$ near $T_c$ (figure~\ref{e22reT}). 
At frequencies $\nu \gg \nu_{s}$, the dielectric constant behaves as a pure lattice contribution. This corresponds to the frequency range $\nu>10^{10}$~Hz on the frequency dependences $\varepsilon_{22} (\nu)$ in figure~\ref{e22re_nu_h},\ref{e22re_nu_hs}.

The increase in the relaxation time $\tau$ and the decrease in the relaxation frequency $\nu_ {s}$ as we approach the temperature $T=T_c$, is manifested in the shift of the  region of dispersion to lower frequencies on the frequency dependence $ \varepsilon_{22} (\nu) $ (figure~\ref{e22re_nu_h},\ref{e22re_nu_hs}), as well as in the expansion of the ``depression'' of $\varepsilon_{22}'$ and of the peak on $\varepsilon_{22}''$ when approaching the temperature $T=T_c$ (figure~\ref{e22reT}). Since  $\nu_{s} \rightarrow 0$ at $T=T_c$, this ``depression'' [as well as the peak on $\varepsilon_ {22}''(T)$] exists at all frequencies; at low frequencies, it is very narrow. The value of permittivity $\varepsilon_{22}'$ at the minimum point (at $T=T_c$) is equal to the lattice contribution $\varepsilon_{22}^{0}$.

\begin{figure}[!h]
	\begin{center}
		\includegraphics[scale=0.85]{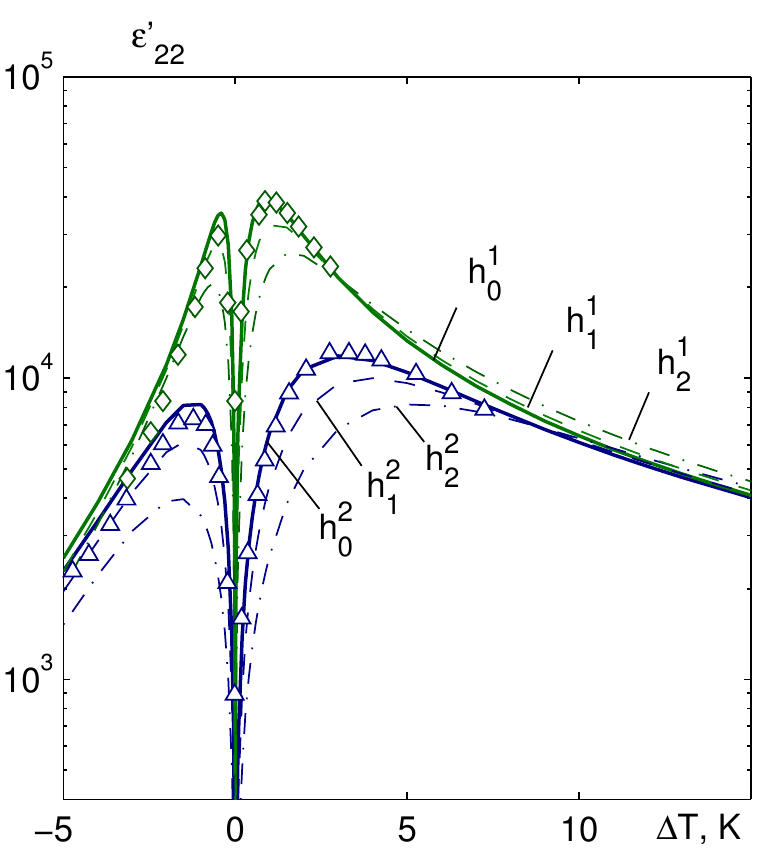}  \includegraphics[scale=0.85]{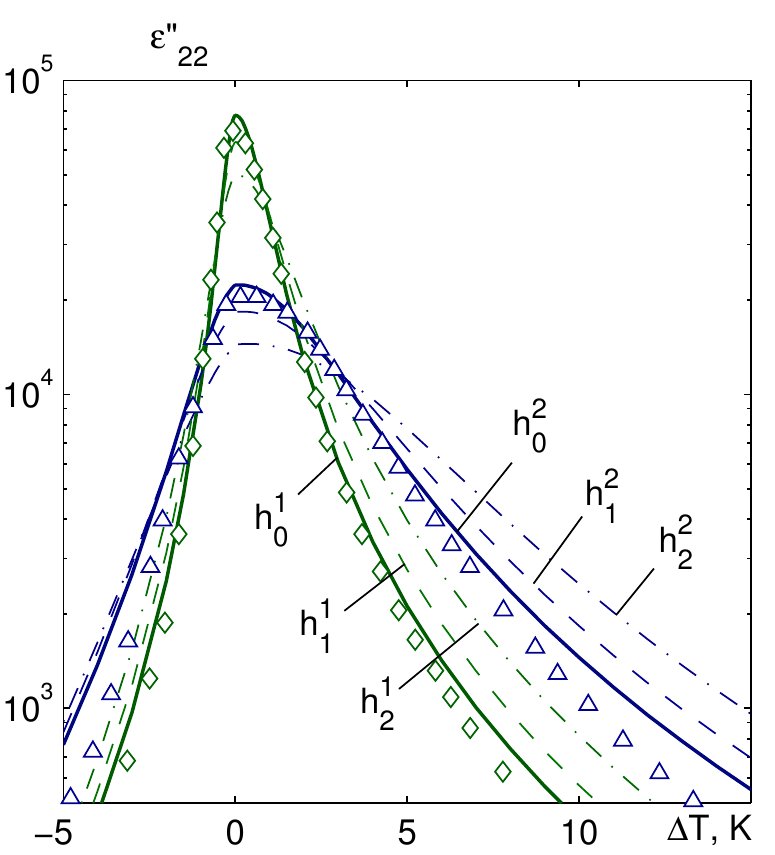}
	\end{center}
	\vspace{-5mm}
	\caption{(Colour online) The temperature dependences of  $\varepsilon'_{22}$ and  $\varepsilon''_{22}$ of CDP
		at different frequencies $\nu$  and at different values of hydrostatic pressure $p_h$. In the notation $h^{i}_p$, the superscripts number the frequency $\nu$ (MHz): 72.4 -- 3, $\diamond$, \cite{Deguchi3575};  251.2 -- 4, $\vartriangle$, \cite{Deguchi3575}, and the lower indices indicate the value of pressure (kbar).} \label{eps22_re_h_34}
\end{figure}

Hydrostatic pressure $p_{h}$ lowers the phase transition temperature $T_c$ and reduces the value of $\tau^{-1}$ at a fixed $\Delta T=T-T_c$, (figure~\ref{tau_p}, curves h$_1$, h$_2$, h$_3$) compared to $\tau^{-1}$ without pressure (curve h$_0$).
As a result, the dispersion region shifts to lower frequencies (figure~\ref{e22re_nu_h},\ref{e22re_nu_hs}, curves h$^i_1$, h$^i_2$), compared to the case of no pressure (curves h$^i_0$).
While on the temperature dependence of the permittivity, the ``depression'' on $\varepsilon_{22}'$ and the peak on $\varepsilon_{22}''$ (figure~\ref{eps22_re_h_34}, curves h$^i_1$, h$^i_2$) broaden under pressure, compared with the case of no pressure (curves h$^i_0$).%

In addition, at low frequencies in the paraelectric phase (figure~\ref{e22re_nu_h}) and at temperatures far from $T_c$ the permittivity $\varepsilon_{22}'$ in figure~\ref{eps22_re_h_34} in the presence of pressure is greater than without pressure, because it is close to static in this area.  The static permittivity at $\Delta T=$ const increases with pressure (figure~\ref{tau_p}b).
In the ferroelectric phase at low temperatures, the pressure reduces the dielectric constant.
\newpage

\section{Conclusions}

It is established that the dynamic dielectric constant of CDP at low frequencies behaves as static; at frequencies commensurate with the
inverse relaxation time, there is a relaxation dispersion; and at high frequencies, only the lattice contribution is manifested
in permittivity. The region of the longitudinal dispersion in CDP shifts to low frequencies when the temperature approaches the phase transition point, which is associated with a significant increase in relaxation time when approaching the temperature $T_{c}$.
A satisfactory agreement of theoretical results with experimental data is obtained.

The effect of hydrostatic pressure on the dielectric properties is manifested in a decrease of the phase transition temperature, in an increase of the static dielectric constant in the paraelectric phase and in an increase of the relaxation time. This leads to an increase of the dynamic permittivity at  pre-relaxation frequencies and to the shift of the dispersion region to lower frequencies. In the ferroelectric phase at low temperatures, the pressure reduces the dielectric constant.
The dynamic permittivity is monodisperse.

As you can see, the dielectric constant in the dispersion region is quite sensitive to the pressure. Thus,
the CDP crystal can be used as a high-frequency filter in which the absorption and transmission spectra can be adjusted by pressure.

\ukrainianpart

\title{Вплив гідростатичного тиску на  динамічні діелектричні характеристики сегнетоелектрика  CsH$_2$PO$_4$}
\author{ А. С. Вдович \refaddr{label1}, Р. Р. Левицький \refaddr{label1}, І. Р. Зачек \refaddr{label2}}
\addresses{
	\addr{label1} Інститут фізики конденсованих систем Національної академії наук України, \\вул. Свєнціцького, 1, 79011 Львів, Україна
	\addr{label2} Національний університет ``Львівська політехніка'', Україна, 79013, Львів, вул.~С.~Бандери,  12}

%
%
%

\makeukrtitle

\begin{abstract}
\tolerance=3000%
На основі псевдоспінової моделі деформованого кристала  CsH$_2$PO$_4$  в рамках методу Глаубера
отримано рівняння для залежного від часу середнього значення псевдоспіна, яке розв'язано у  випадку малих відхилень від стану рівноваги.
Використовуючи розв'язок  рівняння знайдено  вирази для поздовжньої динамічної діелектричної проникності та часу релаксації.
На основі запропонованих параметрів теорії розраховано і досліджено температурну і частотну залежність  динамічної діелектричної проникності та температурну залежність  часу релаксації. Проведено детальний числовий аналіз отриманих результатів. Вивчено вплив гідростатичного тиску  на динамічні характеристики  CsH$_2$PO$_4$.
\keywords сегнетоелектрики, сегнетоелектричний фазовий перехід, проникність, механічні деформації.
	
\end{abstract}

\end{document}